\documentclass[aps,prb,reprint,groupedaddress,showpacs]{revtex4}

\usepackage{graphicx}
\usepackage{bm}        
\usepackage{amsmath}   
\usepackage{epstopdf}

\begin{document}


\title{Pressure-dependent transition from atoms to nanoparticles in magnetron sputtering: Effect on WSi$_2$ film roughness and stress}


\author{Lan Zhou}
\author{Yiping Wang}
\altaffiliation[Present address: ]{Nanjing University of Aeronautics and Astronautics, Nanjing, China.}
\author{Hua Zhou}
\altaffiliation[Present address: ]{Chemical Sciences and Engineering Division, Argonne National Laboratory, Argonne, Illinois 60439.}
\author{Minghao Li}
\author{Randall L. Headrick}
\email[]{rheadrick@uvm.edu}
\affiliation{Department of Physics and Materials Science Program, University of Vermont, Burlington, Vermont 05405, USA}
\author{Kimberly MacArthur, Bing Shi}\author{Ray Conley}
\altaffiliation[Present address: ]{National Synchrotron Light Source II, Brookhaven National Laboratory, Upton, New York 11973.}
\author{Albert T. Macrander}
\affiliation{Advanced Photon Source, Argonne National Laboratory, Argonne, Illinois 60439, USA}


\date{\today}

\begin{abstract}
We report on the transition between two regimes from several-atom clusters to much larger nanoparticles in Ar magnetron sputter deposition of WSi$_2$, and the effect of nanoparticles on the properties of amorphous thin films and multilayers. Sputter deposition of thin films is monitored by $\emph{in situ}$ x-ray scattering, including x-ray reflectivity and grazing incidence small angle x-ray scattering. The results show an abrupt transition at an Ar background pressure $P_c$; the transition is associated with the threshold for energetic particle thermalization, which is known to  scale as the product of the Ar pressure and the working distance between the magnetron source and the substrate surface. Below $P_c$ smooth films are produced, while above $P_c$ roughness increases abruptly, consistent with a model in which particles aggregate in the deposition flux before reaching the growth surface. The results from WSi$_2$ films are correlated with $\emph{in situ}$ measurement of stress in WSi$_2$/Si multilayers, which exhibits a corresponding transition from compressive to tensile stress at $P_c$. The tensile stress is attributed to coalescence of nanoparticles and the elimination of nano-voids.
\end{abstract}

\pacs{61.05.cf, 68.35.Ct,  68.55.A-, 68.60.Bs , 68.65.Ac, 81.10.Bk, 81.15.Aa, 81.15.Cd, 82.30.Nr}

\maketitle



\section{INTRODUCTION}
Magnetron sputtering has been widely used since the 1970s as a deposition method for metal, semiconductor, and other inorganic thin films for applications such as optical coatings and microelectronic circuits. \cite{Freund2003} Recently, there has been renewed interest in applications that require strict control over thin film structure and mechanical properties. For example, multilayers fabricated by sputter deposition are in development for x-ray optics such as in multilayer Laue lenses for nanometer scale focusing of x-rays. \cite{macrander2000, Liu2005, Kang2006,Kang2008,Conley2008} They require sub-nanometer roughness over several thousand layers with minimal built-in stress, and these specifications  have proven to be difficult to achieve with state of the art deposition techniques. Varying the background gas pressure can produce significant effects in thin films and multilayers, such as an abrupt change in interface roughness, \cite{Fullerton1993} or a sudden transition in film stress. \cite{Freund2003, Nix1999} For example, Fullerton $\emph{et al.}$\cite{Fullerton1993} have found that interfacial roughness in Nb/Si multilayers increases dramatically when the Ar pressure exceeds 9 mTorr. Cyrille $\emph{et al.}$\cite{Cyrille2000} have observed a similar effect in Fe/Cr multilayers, and have made use of the pressure-induced roughness to enhance the magetoresistance in these structures.  Similarly, striking changes in film stress have been observed for a number of materials. Hoffman and Thornton\cite{Hoffman1977} have found that a stress transition from compressive to tensile occurs at a pressure that depends on the atomic mass of the deposited material,  and that the transition pressure is inversely proportional to the atomic mass of the sputtering gas. \cite{Hoffman1980} It would be beneficial to understand the origin of these effects because it would potentially lead to improved processes for fabrication of films for a variety of applications, and because there is significant interest in understanding the fundamental mechanisms that govern surface dynamics during film deposition.

In this paper, we demonstrate that the sudden changes in roughness and stress as the background gas pressure is varied both arise from an effect that is intrinsic to the magnetron sputtering process. There is a transition in the sputtered flux in which the dominant species produced by the magnetron source abruptly changes from atomic-size species to nanoparticles containing several hundred atoms. Our analysis methods involve $\emph{in situ}$ film deposition studies using synchrotron x-ray scattering to observe cluster distributions on the substrate surface before film coalescence and the power spectrum of surface roughness of films deposited from atoms vs from nanoparticles. The spectra are used to extract parameters related to the roughening processes on the film surface using standard equations of x-ray scattering.\cite{Sinha1988} The results show that surface roughening of deposited films at different pressures increases dramatically above a critical background pressure $P_c$. Analysis of the data shows that the increase in roughness cannot be explained by surface relaxation processes alone. Rather, it is caused by a sharp increase in the deposited particle volume, which results in increased roughness because larger particles contribute more deposition noise.\cite{Moseler2000, Barabasi1995} Specifically, we have investigated amorphous WSi$_2$ film deposition using $\emph{in situ}$ synchrotron x-ray scattering and infer that there is a transition from atoms to nanoparticles in magnetron sputtering when the sputtering pressure is raised above $P_c$ = 6 mTorr.  These results are described in detail in Sec. \ref{thinfilmresults}. The data analysis are described in Sec. \ref{sec:theorybackground}.

We also report a complementary set of experiments involving measurement of thin film stress during the deposition of WSi$_2$/Si multilayers. The results confirm that the stress transition occurs at $P_c$ = 6 mTorr, in good agreement with the results of the film roughness experiments. The tensile stress observed above the transition is interpreted as arising from coalescence of deposited clusters. The effect is similar to reports in the literature of tensile stress arising from coalescence of hills formed on the growth surface of amorphous films in the later stages of growth due to continuous viscous coalescence, and driven by surface tension due to the large curvature at the cusps between the hills. \cite{Mayr2001} This idea becomes more compelling with the realization that the deposition flux above the stress transition is fundamentally composed of nanoclusters, which may significantly contribute to tensile stress as they coalesce into a continuous film through the elimination of nano-voids. These results are presented in Sec. \ref{multilayerresults}.

This observation of a particle volume transition opens up  possibilities for fundamental studies of nanoparticle aggregation, and  applications such as film deposition and crystal growth via nanoparticle assembly. Production of clusters containing hundreds of atoms by aggregation of vapor has previously been demonstrated only in specialized gas aggregation instruments operating in a higher pressure range not typically used for sputter deposition.\cite{Dietz1981, Sattler1980} Our observations are for one specific material, however it is clear that the effect extends to a variety of other materials\cite{Hoffman1977} and possibly also to other deposition techniques such as pulsed laser deposition (PLD). In this regard, we note that PLD has some relevant similarities to magnetron sputtering, particularly the production of a dense vapor of energetic particles, which might  produce clusters through the mechanisms of vapor-phase aggregation that we discuss in this paper. Furthermore,  linking the stress transition and roughening transition to the particle volume suggests  new possibilities for tailoring structural and mechanical properties of thin films and multilayers.

\section{EXPERIMENTAL DETAILS}
\subsection{WSi$_2$ thin film deposition and $\emph{in situ}$ x-ray scattering}
\begin{figure}[htbp]
	\includegraphics[width=5.0 in]{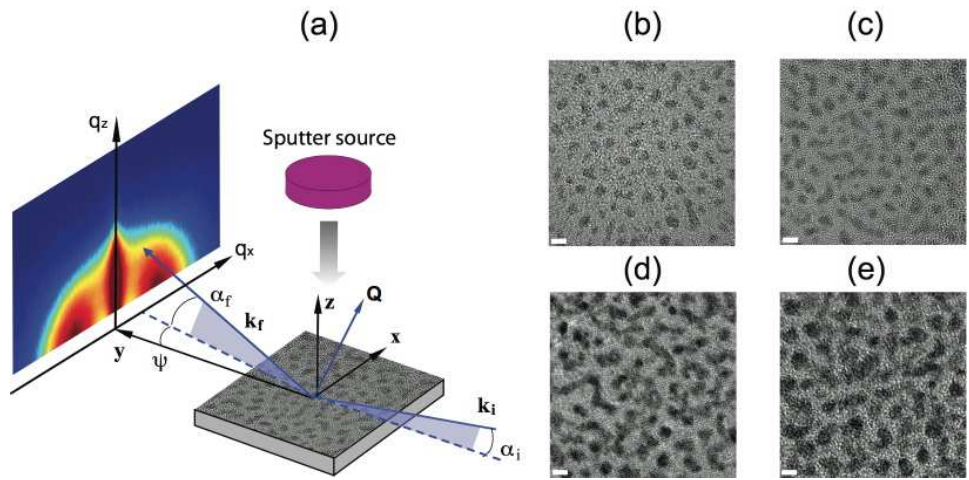}
	\caption{(a) Schematic of real-time grazing small-angle x-ray scattering (GISAXS) measurements during sputter deposition. The image is shown rotated with respect to the actual experiment, where the $x$ direction is vertical. The incident x-ray beam (wave vector $\bm{k_i}$) impinges on the sample surface under a grazing incidence $\alpha_i$. The scattered intensity is recorded as a function of the out-of-plane angle $\alpha_f$ and the in-plane angle $\psi$. The wave vector transfer is denoted by $\bm{Q}=\bm{k_f}-\bm{k_i}$.  A 2D GISAXS reciprocal-space map is shown, which is acquired by varying $\alpha_f$ through the range $0.6^\circ$--$10^\circ$ while keeping $\alpha_i = 0.2^\circ$. This scan is repeated at two detector positions to cover $\psi$ angle ranging from $-0.2^\circ$ to $5.56^\circ$.  The negative $q_x$ part is mirror imaged. This $\alpha_f$ scan closely approximates a $q_z$ vs $q_x$ map of the diffuse scattering component since $q_y$ is very small in this GISAXS geometry.  For simplicity of notation we drop the subscript on $q_x$ in reporting the results in the main text since the surfaces are assumed to be isotropic.   The illustration also shows  a TEM image of WSi$_2$ film of nominal thickness 1 nm for a visual representation the sample surface in real space.  [(b)--(e)] TEM images of a series of WSi$_2$ films deposited on thin carbon grids at 8 mTorr at room temperature. The nominal film thicknesses are 0.5 nm, 1.0 nm, 2.0 nm, and 3.0 nm, respectively. The small white bar at the lower left of each image  indicates a length scale of 2 nm.   \label{fg:gisaxsandtem}}
\end{figure}

The growth experiments were performed in a custom-built ultrahigh vacuum chamber with a base pressure of $10^{-10}$ Torr installed at X21 station of National Synchrotron Light Source (NSLS), Brookhaven National Laboratory (BNL). A schematic of the experiment is shown in Fig. \ref{fg:gisaxsandtem}(a). We take the $z$ direction to be sample normal and the $y$-direction to be along the projection of the incident x-ray beams onto the sample surface. The $x$ direction is vertical and the $y$-$z$ scattering plane is horizontal in the laboratory frame. All amorphous WSi$_2$ samples were prepared at room temperature by a dual-gun dc magnetron sputtering system (2" Meivac MAK) using ultrahigh purity (99.999\%) Ar gas. A water-cooled 2-inch-diameter WSi$_2$ target (purity 99.999\%) was mounted on the gun, which was 90 mm away from the sample surface. The target normal lies in $x$-$z$ plane  and 7$^\circ$ to the substrate surface normal ($z$ direction). Before deposition the target was pre-sputtered for 5 min with the target shutter closed. The dc sputtering power was kept constant at 50 W and the Ar pressure was adjusted ranging from 3 to 18 mTorr. The pressure was monitored by Pirani gauge (Stanford Research System), which was calibrated for Ar gas. The pressure varied by less than 0.1 mTorr throughout the deposition process. Under these conditions, the resulting deposition rate $v$ was between 0.103 and 0.111 nm/s at all pressures. Amorphous thermal SiO$_2$/Si with rms roughness $\sigma_{rms}$ = 0.25 nm were used as substrates for x-ray scattering measurements.

The synchrotron x-ray flux at the X21 beamline is approximately $1\times10^{12}$ photons/s at photon energy of 10 keV (wavelength $\lambda$ = 0.124 nm$^{-1}$) after the Si(111) double-crystal monochromator and toroidal mirror. A slit of dimension 0.2$\times$0.5 mm$^2$ (vertical$\times$horizontal) 1070 mm in front of the sample and a slit (50$\times$1 mm$^2$) 790 mm behind the sample define the angle of incidence $\alpha_i$ and the exit angle $\alpha_f$, respectively. The linear detector is oriented parallel to the sample surface. It consists of 384 pixels with 8 pixels/mm along the $x$ direction. It covers a range of in-plane angle $\psi=2.88^\circ$ at a distance of 955 mm behind the sample. The ultimate resolution of the instrument within the scattering plane is given by $\delta q_z\approx 0.09$ nm$^{-1}$ for the direction perpendicular and $\delta q_y\approx10^{-3}\times q_z$ nm$^{-1}$ parallel to the surface, while the resolution perpendicular to the scattering plane $\delta q_x\approx 0.007$ nm$^{-1}$.

Four types of scans were performed during or after each deposition:  (i) real-time grazing incidence small-angle x-ray scattering (GISAXS) monitoring of the evolution of surface morphology. The scattered intensity was measured by a one-dimensional detector oriented parallel to the sample surface.  The exit angle $\alpha_f$ was kept constant at 0.6$^\circ$  and  the incident angle was fixed at  $\alpha_i =$ 0.2$^\circ$ to the sample surface.  This asymmetric scattering geometry avoids the saturation of the linear detector because the specular beam does not reach the detector. In addition, the angle of incidence is  below the WSi$_2$ critical angle for total external reflection at 10 keV photon energy in order to enhance the surface sensitivity, and to keep $q_z$ as low as possible. Each spectrum thus represents the scattered intensity as a function of $q_x$ on a range of momentum transfer 0.003 nm$^{-1}\le q_{x} \le$ 2.5 nm$^{-1}$ at constant  $q_z = 0.7$ nm$^{-1}$.  This gives access to the surface roughness on lateral length scales between a few nanometers and 1 $\mu$m.   (ii) $\emph{In situ}$ specular reflectivity scans after each deposited layer. This was done by rotating both the detector and sample about the $x$ axis to keep the angle $\alpha_i = \alpha_f$. The intensity was recorded by the sum of central 8 pixels of the linear detector. The surface roughness and thickness of these films were derived from those scans using a least-squares fit in which the reflected intensity was computed based on recursive application of the Fresnel equations;\cite{Windt1998} (iii) $\emph{in situ}$ $\psi$ scans after each deposited layer. This scan circumvents the limitation of the length of the linear detector, which determines the range of $q_x$ accessible in GISAXS with a fixed detector position. The linear detector is rotated about the sample normal by exactly the length of the detector at each step of the scan.  For this study, we have employed scans to $q_x \approx 10$ nm$^{-1}$. This extended $q_x$ range is particularly useful for quantitative modeling of the roughness spectrum.  (iv) Two-dimensional GISAXS $\psi$~ -~ $\alpha_f$ maps acquired while keeping $\alpha_i$ fixed. This $\alpha_f$ scan closely approximates a $q_z$ vs $q_x$ reciprocal map of the diffuse scattering component since $q_y$ is very small in the GISAXS geometry. This type of scan is shown in Fig. \ref{fg:gisaxsandtem}(a).

\subsection{Transmission electron microscopy (TEM) study of WSi$_2$ cluster coalescence\label{TEMexperimentaldetails}}

Amorphous ultrathin carbon coated copper grids (PELCO, No. 01824 from Ted Pella, Inc.) were used as substrates for TEM observations. TEM images were obtained on a JOEL 2100F high-resolution analytical transmission electron microscope operating at 200 kV at the Center of Functional Nanomaterials (CFN) at BNL.  WSi$_2$ clusters deposited on these carbon surfaces and observed via TEM were found to be in good agreement with cluster distributions observed via x-ray scattering on SiO$_2$ surfaces.  Both carbon and SiO$_2$ are relatively inert, and therefore would not be expected to produce  dramatically different results for our study, where surface diffusion and particle migration play a minor role.

\subsection{WSi$_2$/Si Multilayer Experiments}
A well known technique to measure stresses in thin films is to measure the curvature and then to apply Stoney's equation,\cite{Stoney1909} which in its biaxial form is  given by \cite{Nix1989, Janssen2009}
\begin{equation}
\frac{1}{R}= 6 \sigma \left(\frac{1-\nu}{E}\right)\frac{t_f}{t_s^2}~~~,
\end{equation}

\noindent Here the curvature is $1/R$, where $R$ is the radius of curvature of a wafer, $t_f$ and $t_s$ are the thickness of the film and substrate, respectively, $\sigma$ is the stress in the film, $E$ is the Young's modulus of the substrate, and $\nu$ is the Poisson's ratio of the substrate. This equation applies to biaxial bending in the thin film limit, that is, in the limit in which the effective elastic constants are those of the substrate. We note that if the stress in the film is constant, then the prefactor in Stoney's equation is constant, and the above equation yields a linear relationship between curvature and film thickness.

For thin films grown on Si substrate wafers, one must account for the fact that the elastic constants of Si are anisotropic in evaluating $E$ and $\nu$.\cite{Brantley1973} However, for bending of Si(100) wafers, the biaxial modulus, $E/(1-\nu)$, is conveniently isotropic in the plane of the wafer.\cite{Brantley1973}

Curvatures in WSi$_2$/Si  multilayers were measured $\emph{in situ}$ in a growth chamber located at Argonne National Laboratory. Substrates were positioned on central vertical rotary axis, and sputtering proceeded horizontally. The chamber was equipped with a multibeam optical sensor (MOS) supplied by kSA.\cite{KSA} The MOS was operated with a laser and beam splitters to produce a 4$\times$3 (column$\times$row) grid of beams reflected from a multilayer sample. The sample curvature was measured in both the horizontal and vertical directions. The MOS was situated on a vacuum port adjacent to the sputtering gun, and curvature measurements were made by periodically bringing the sample from a position facing the sputtering gun to a position facing the MOS. This was done at intervals corresponding to 1.1 nm of thickness.

The substrates were 50 mm diameter Si(100) wafers. Multilayers consisting of 20 bilayers of WSi$_2$/Si with layer thicknesses of 5.5 nm each. Multilayers were studied for the Ar plasma pressures of 2.3, 6, 12 and 18 mTorr. These parameters were chosen to be the same as for an earlier $\emph{in situ}$ x-ray reflectivity study,\cite{Wang2007} and layer thicknesses were checked subsequently with $\emph{ex-situ}$ x-ray reflectivity measurements made after a full multilayer was grown.

\section{RESULTS of WSi$_2$ thin film deposition experiments}\label{sec:resultsanddiscussion}

\subsection{Transient stage of WSi$_2$ deposition\label{TEMresults}}

Figure \ref{fg:gisaxsandtem}(a) displays a 2D GISAXS map of WSi$_2$ clusters deposited on thermal SiO$_2$/Si substrate at 8 mTorr for 10 s, which has a nominal film thickness of 1 nm. Symmetrical rounded shapes of the scattered intensity are observed, indicating the presence of  three-dimensional clusters on the growing surface. The peak of the intensity occurs at $q_{peak} \approx 2 $ nm$^{-1}$, indicating a particle separation of $2 \pi / q_{peak} \approx $ 3 nm, and the width of the peaks indicates that the cluster positions are nearly randomly distributed.

Figure \ref{fg:gisaxsandtem}(b)-(e) shows TEM images of WSi$_2$ clusters on amorphous carbon grids with different nominal film thicknesses from 0.5 to 3 nm. They were deposited at 8 mTorr with deposition rate of 0.105 nm/s. Clusters 1 - 2 nm in diameter are seen at 5 s, which corresponds to about 0.5 nm average thickness.  At later times the clusters aggregate into elongated meandering islands on the surface, indicating limited coalescence. Similar clusters are also observed at 4 mTorr (not shown in the figure).  The TEM observations are consistent with the GISAXS pattern.

These observations clearly show that clusters are present on the surface during the early stage of film deposition before coalescence. One of the key questions that we endeavor to answer in this study is whether the clusters form in the sputtering plasma before reaching the substrate, or whether the aggregation occurs on the surface. Both possibilities are plausible. Aggregation in the gas phase depends on  collisions of energetic particles with gas molecules, which produces a thermalization effect. This process is expected to be pressure dependent (for reference, we note that the mean-free path in Ar at 8 mTorr is about 6.8 mm). On the other hand, surface aggregation depends on the mobility of particles after they land on the surface. Processes such as thermal diffusion should be very limited when the substrate is held at room temperature, and so the length scale of aggregation in this case is expected to be short.  More generally, both gas aggregation $and$ surface aggregation processes may act in concert to produce the final structures observed. We believe that this is the most likely explanation of the fact that we observe clusters on the surface at all pressures above and below $P_c$. Therefore,  the observed cluster size of 1 - 2 nm observed at the earliest stage  of deposition should be taken as an upper limit of the cluster size deposited from the magnetron source. We will return to the question of deposited particle size and surface cluster size in the next section, and also in the data analysis presented in Sec. \ref{sec:theorybackground}.

\subsection{Pressure dependence and temporal evolution of surface roughness\label{thinfilmresults}}

\begin{figure}[htbp]
 	\includegraphics[width=3 in]{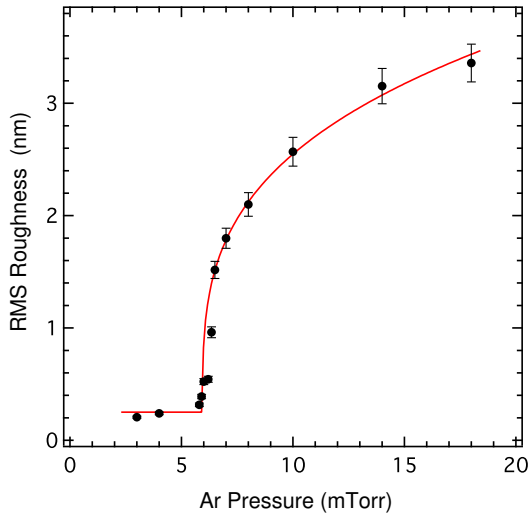}
 	\caption{The rms roughness of $t_f = 180$-nm-thick WSi$_2$ films as a function of Ar background pressure in which two roughness regimes are observed. The red line is a power law $(P-P_c)^s$.   \label{fg:roughnessvspressure}}
\end{figure}

The rms roughness derived from specular x-ray reflectivity is plotted in Fig. \ref{fg:roughnessvspressure} as a function of Ar pressure. The films deposited at different Ar pressures each have the same thickness $t_f =$ 180 nm. The surface roughness shows a very abrupt transition at 6 mTorr where the roughness increases dramatically. It is well described by a power law $(P-P_c)^s$, with $P_c$ = 6 mTorr and $s = 0.30 \pm 0.05$.

A striking dependence of surface morphology of thin films and multilayers on the sputtering gas pressure has been observed previously.\cite{Stearns1991, Fullerton1993, Cyrille2000, Kim2006}  At low gas pressures, sputtered and back-reflected species ballistically impinge on the growing surface with higher average kinetic energies, which leads to  enhanced surface relaxation and smooth surfaces.  Surface smoothing by energetic particle bombardment at low pressures is consistent with our data, as we discuss below; see  Fig. \ref{fg:smoothingrealtime} and the discussion of mechanisms in Sec.  \ref{sec:PGCdiscussion}. The number of gas-phase collisions in the path from target to substrate increases with  pressure so that above a "thermalization pressure", the velocities of energetic particle emerging from the source are reduced to thermal velocities before they strike the substrate.\cite{Somekh1984, SomekhVacuum1984}    Our suggestion for explaining the roughening transition, which we discuss further in Sec.  \ref{sec:PGCdiscussion}, is to combine this thermalization effect with a model of cluster growth in the gas phase.  This model predicts a very sharp onset of cluster growth and hence roughening  as the sputtering pressure is varied, in agreement with the results shown in Fig.  \ref{fg:roughnessvspressure}.

Before proceeding, we mention several alternative models for the roughening transition based on changes in the strength of smoothening processes with constant particle size: (i)  a possible model is that the surface roughens above the thermalization pressure simply because energetic smoothing processes are suppressed above $P_c$. This model seems highly plausible.  However,  this model  is inconsistent with the shape of the curve in Fig.  \ref{fg:roughnessvspressure}. In particular,  a gradual variation in the strength of the  smoothing processes as a function of  the background Ar pressure is predicted, which would produce a continuous increase in the roughness. In contrast, we observe  a sudden onset at $P_c$. (ii) A second possibility is a transition from stable surface kinetics at low pressures to an unstable surface at higher pressures, which can be understood in terms of a change in sign of one of the smoothing coefficients in a linear continuum model.\cite{Zhou2008}  A  sharp transition near $P_c$ is predicted as a surface instability mechanism becomes dominant.\cite{Sigmund1969, Mayer1994, Carter2001, Bradley1988} In this model, the surface would transition to an unstable surface that would be characterized by rapidly increasing amplitude at a preferred spatial wavelength. In contrast, as we describe below, the surface is found to exhibit stable kinetic roughening in both pressure regimes  below \emph{and above} $P_c$.  Therefore, \emph{alternative models for the roughening transition  based on changes in the strength of smoothening processes alone with no variation in particle size are inconsistent with the data presented.}

\begin{figure}[htbp]
	\includegraphics[width=4.5 in]{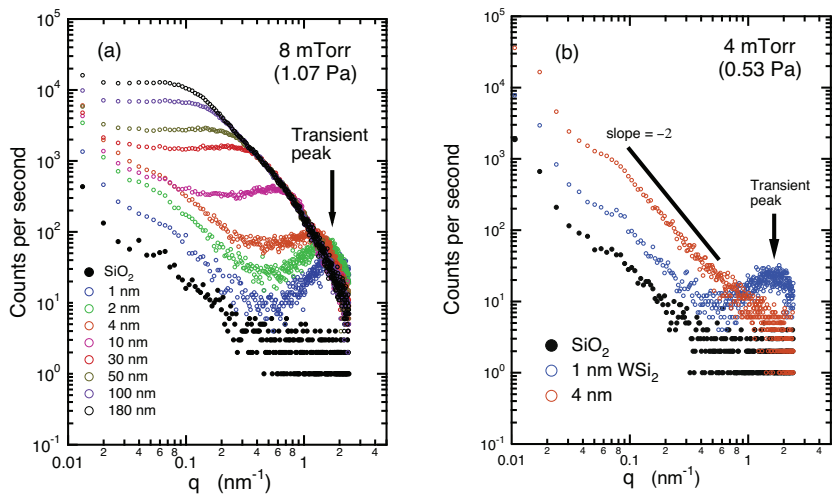}
	\caption{Real-time GISAXS evolution during sputter deposition (a) at 8 mTorr and(b) at 4 mTorr. The arrows indicate the transient peaks observed in the precoalescence stage.  \label{fg:psdvstimeattwopresures}}
\end{figure}

Figure \ref{fg:psdvstimeattwopresures} shows real-time GISAXS evolution during the deposition at 8 and 4 mTorr, respectively. In both cases, with growing amount of WSi$_2$, a transient peak associated with 3D clusters first appears and moves towards lower $q$, indicating coalescence.  This is consistent with the TEM results in Fig. \ref{fg:gisaxsandtem}(b). Film coalescence is complete at about 4 nm in each case, which begins the kinetic roughening stage of the surface evolution.  Below, we discuss Fig.  \ref{fg:psdvstimeattwopresures}(a) and \ref{fg:psdvstimeattwopresures}(b) in turn.

Figure \ref{fg:psdvstimeattwopresures}(a) illustrates the typical evolution of the GISAXS intensity for pressures above $P_c$. After the film coalescence the diffuse scattering profile $S(q)$ increases monotonically with film thickness for $q < q_c$, where $q_c$ is the cutoff wave number. As detailed in Sec. \ref{sec:dynamics}, after a  correction  at low $q$,  the GISAXS data are proportional  to the power spectral density (PSD)  which is expected to be constant  for $q < q_c$, i.e., the spectrum is expected to be white below  a  cutoff wave number.  We note that $q_c$ is related to the surface correlation length $\xi$, the maximum length scale to which the surface roughness has propagated,  by $q_c = 2\pi / \xi$. The intensity increases (at least) linearly with time since $S(q) \propto \Omega~ t_f^{(2+2\alpha)/z}$ for $q<q_c$, where $\Omega$ is the particle volume and $t_f$ is the film thickness, and typical values for the exponent $(2+2\alpha)/z$ range from 1.0 for linear models such as the Edwards-Wilkinson  (EW)  model\cite{EW1982}  to $\sim 1.75$ for the Kardar-Parisi-Zhang  model.\cite{Kardar1986} See Sec.  \ref{sec:theorybackground} for further discussion of the scattering intensities.  We interpret this rapid increase in the intensity to be indicative of a large particle volume.  The spectrum also exhibits characteristics of kinetic roughening because the curves for different film thickness overlap perfectly at $q > q_c$,\cite{Stearns1993, Barabasi1995} which also rules out the possibility that the surface is unstable. At later times, a crossover between scaling regimes exhibiting different exponents  is observed at $\sim$ 1 nm$^{-1}$. We interpret this as being due to two distinct relaxation processes for length scales larger and smaller than this characteristic length because of the change in the slope of the curve above and below 1 nm$^{-1}$. Note that this feature does not shift after the coalescence, which suggests that it is the result of steady-state processes rather than being simply a remnant of the transient feature.

Figure \ref{fg:psdvstimeattwopresures}(b) illustrates the behavior below $P_c$.  After coalescence, the film surface is statistically similar to that of the starting surface.  This is  because the shot noise is very low due to the smaller particle volume, making the increase for $q<q_c$ and the cutoff itself very difficult to observe at low sputtering pressures. At later times the spectrum  is nearly identical to the one shown for $t_f = 4$ nm even for much thicker films. The  $q^{-2}$ dependence of the curve is the stable spectrum, indicating an EW-type smoothing mechanism. The roughening transition is therefore interpreted as a transition from an EW regime at low pressures to a new kinetic roughening regime above $P_c$ that is triggered by the larger particle volume.

\begin{figure}[htbp]
	\includegraphics[width=6.0 in]{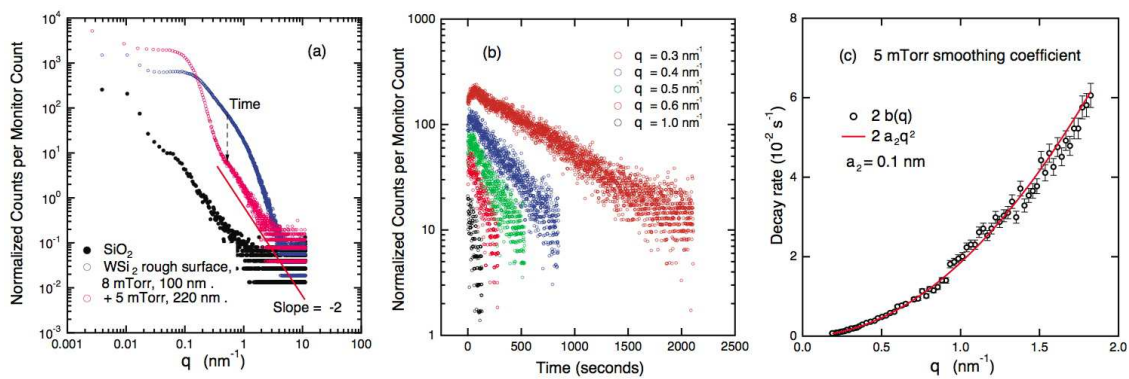}
	\caption{(a) GISAXS spectra show the PSD profiles after deposition of the smoothing layer at 5 mTorr (red) on rough layer deposited at 8 mTorr (blue). The substrate spectrum is also shown for comparison. During the deposition of the low-pressure layer, the roughness decreases from 1.6 nm to 0.66 nm. The smaller wavelength surface roughness decays away faster and the corresponding roughness spectrum exhibits a power law with an exponent of $-2$; (b) GISAXS intensity evolution at various $q$ in a semilog plot. The high spatial frequency features vanish faster. The slope of the curves shows the decay rate, which is plotted in (c). (c)Plot of the decay rates, confirming that the relaxation is dominated by a $q^2$ term, as indicative of the EW model. The $a_2$ relaxation coefficient is also extracted from this curve.\label{fg:smoothingrealtime}}
\end{figure}

In Fig. \ref{fg:smoothingrealtime}, an experiment to determine the smoothing coefficients is shown, which consists of depositing a relatively rough 100 nm film at 8 mTorr,  and then smoothing the surface by depositing 220 nm at 5 mTorr.  This data can be used to extract surface relaxation coefficients directly from the smoothing kinetics. We have previously used this technique to study relaxation mechanisms in ion bombardment induced smoothing of surface.\cite{Zhou2008} The corresponding GISAXS profiles for SiO$_2$ substrate, rough and smoothed WSi$_2$ layers are shown in Fig. \ref{fg:smoothingrealtime}(a). It can be seen that the diffuse intensity decreases during deposition of WSi$_2$ at 5 mTorr, consistent with the decrease in the rms surface roughness, which was determined by specular x-ray reflectivity. Note that  the increasing intensity for $q<0.1$ nm$^{-1}$ is due to the fact that the initial GISAXS curve for the rough layer does not accurately give the spectrum of roughness because the small roughness condition is not satisfied due to the large initial surface roughness  (see Sec. \ref{SRAbreak}). The spectrum of the smooth layer exhibits a dropoff at $q_c$, the cutoff frequency, which is related to the characteristic correlation length on the surface corresponding to the largest surface features which  have already been smoothed out. Above this spatial frequency, a scaling behavior with an exponent of -2 is observed. This part of the spectrum is identical to the one obtained for WSi$_2$ deposited on smooth SiO$_2$ at pressure below the transition at $P_c$, as shown in Fig. \ref{fg:psdvstimeattwopresures}(b). This behavior confirms  that the stable roughness spectrum is independent of the starting surface.

Figure \ref{fg:smoothingrealtime}(b) shows the same GISAXS intensity plotted vs time at various length scales. The curves follow a decaying exponential form, in good agreement with the prediction of linear theory (Sec. \ref{sec:dynamics}).  The roughness decay rate at different wavelength is extracted by an exponential decay fit for all the curves. The decay rates are plotted as a function of spatial frequency in Fig. \ref{fg:smoothingrealtime}(c). It  shows unambiguously that a $q^2$-dependent smoothing mechanism is dominant on the surface during the deposition below the transition pressure. Thus,  EW-type behavior is confirmed for pressure below $P_c$.

\begin{figure}[htbp]
	\includegraphics[width=2.5 in]{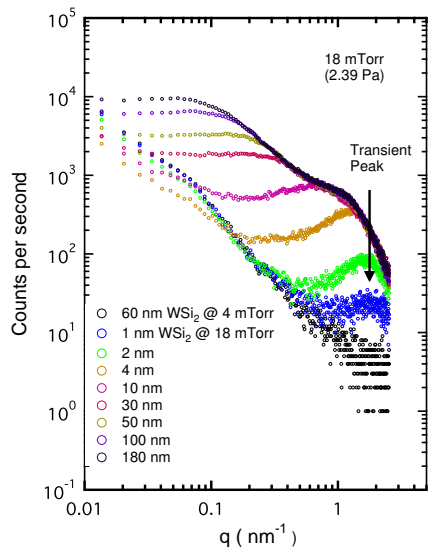}
	\caption{Real-time GISAXS evolution during sputter deposition at 18 mTorr on a smooth layer deposited at 4 mTorr. The arrow indicates the transient peak associated with 3D clusters on the film surface. \label{fg:18on4}}
\end{figure}

Figure \ref{fg:18on4} shows a different variation in the experiment. Here the first layer is deposited at 4 mTorr, producing a very smooth WSi$_2$ surface, then a second layer is deposited at 18 mTorr.  There is no driving force for clustering on the surface when depositing WSi$_2$  onto itself, such as a difference in surface energies. \cite{Steigerwald1988PRL, Steigerwald1988} Therefore, by observing the real-time GISAXS spectrum we can study the transition from smooth to rough without the complication factor of depositing onto a surface that promotes clustering. The results show that the transient peak associated with nanoclusters on the surface appears again at $\approx 2$ nm$^{-1}$.  Note that the transient peak is not part of the kinetic roughening, but rather is due to the deposition of isolated clusters on the substrate.  This is the stage of deposition before the particles begin to interact, and was discussed by Edwards and Wilkinson.\cite{EW1982} This observation clearly confirms that the clusters observed at $P>P_c$ are generated by the sputtering source rather than by aggregation on the surface.

The bump at 1 nm$^{-1}$ for $t_f > 30$ nm in the steady-state spectrum may be related to a  surface instability  process. However, the surface still exhibits kinetic roughening because smoothing processes dominate over the instability during the deposition. The discussion of roughness dynamics is in Sec. \ref{sec:theorybackground}.

\section{THEORY BACKGROUND and analysis of roughening dynamics \label{sec:theorybackground} }

\subsection{Surface morphology dynamics\label{sec:dynamics}}

Combining the surface relaxation mechanisms with the stochastic roughening within a linear approximation, we can write a kinetic rate equation of the Langevin type, for surface growth:

\begin{align}\label{eq:FTLangevin}
 &\frac{\partial{h(\bm{q},t)}}{\partial{t}} =  -b(q)\cdot h(\bm{q},t)+\eta(\bm{q},t)~,~~\\
\label{eq:bq}
 &b(q)=v\sum_{n=2}^4 a_n\cdot q^n~,
\end{align}
	
\noindent where $h(\bm{q},t)$ is the Fourier transform of surface height $h(\bm{r},t)$, $\bm{q}=(q_x,q_y)$ is the in-plane wave number and $q=q_{||}=\left|\bm{q}\right|$ is the circular averaged wave number. The factor $v$ is the deposition rate. The term $b(q)$ on the right of Eq. (\ref{eq:FTLangevin}) models the surface relaxation processes during film growth. Coefficient $a_n$ is a constant characteristic of the specific lateral mass transport mechanism indicated by $n$.  The  term $\eta(\bm{q},t)$ is the reciprocal-space stochastic noise term that describes the random arrival of the depositing species. It has the property

\begin{equation}
\label{eq:noise}
\left<\eta(\bm{q},t)\right> = 0~~, ~\left<\eta(\bm{q},t)\eta(\bm{q}^\prime,t^\prime)\right> =v \Omega~ \delta(\bm{q}+\bm{q}^\prime)\delta(t-t^\prime)~~,
\end{equation}

\noindent which represents uncorrelated  deposition noise with strength $v\Omega$. The factor $v$ is the deposition rate, and $\Omega$ is the volume of the species being deposited, whether they are atoms, molecules, clusters, or nanoparticles.

The PSD of a growing surface is defined as $\left<|h(q,t)|^2\right>$, which is also the Fourier transform of the correlation function $C(r,t)=\left<h(r,t)h(r',t)\right>$. The radially averaged PSD can be analytically determined from Eq. (\ref{eq:FTLangevin}),
\begin{align}
\label{eq:PSD}
&PSD(q,t)=\left<|h(q,t)|^2\right>= PSD(q,0) e^{-2b(q) t}+ v\Omega \frac{ 1 - e^{-2b(q) t}}{2 b(q)} ~~,\\
\label{eq:PSDa}
&PSD(q,t)= \left[PSD(q,0)-PSD(q,\infty)\right] e^{-2b(q) t}+ PSD(q,\infty)~~,~~ \\
\label{eq:PSDinfty}
&PSD(q,\infty)=\frac {v\Omega}{2 b(q)}
\end{align}

\noindent where $PSD(q,0)$ is the power spectral density of the substrate, which equals to zero for perfectly smooth surface. The first term on the right of Eq. (\ref{eq:PSD}) damps the contribution from the substrate, while the second term presents the roughness which increases from intrinsic growth processes. Stochastic roughening by random deposition creates a surface that contains features of all sizes. The PSD is just a constant line for $q < q_c$, i.e., the $q$ spectrum is white. The constant value depends on the rms roughness and increases as $\Omega~ t_f$ for any linear model of the type represented by Eq. (\ref{eq:FTLangevin}). $\Omega$ is the particle volume in deposition flux and $t_f$ is the film thickness.  Eq. (\ref{eq:PSDinfty}) shows in the long-time limit ($t \rightarrow \infty$), the spectrum reaches the steady state that decreases as a power law $q^{-n}$, or multiple power laws if there is more than one nonzero $a_n$,  depending explicitly on the identity of the relaxation mechanisms. In the framework of kinetic roughening, the surface morphology exhibits asymptotic scaling behavior at long times and long length scales. In the case of Family-Vicsek scaling,\cite{Family1985}   the PSD can be expressed by the relation \cite{Barabasi1995}

\begin{equation}
PSD(q,t) \sim  q^{-(2\alpha+2)}f(q/q_c)~~,~~~~ q_c \sim t^{-1/z}
\end{equation}
\noindent with the scaling function

\begin{equation*}
f(u) \sim
\begin{cases}
const, & \text{if } u \gg 1,\\
u^{2\alpha+2} & \text{if } u \ll 1.
\end{cases}
\end{equation*}

\noindent where $\alpha$ is called the roughness exponent, $z$ the dynamic exponent, and $\beta=\alpha/z$ the growth exponent. The integral of $PSD(q)$ over $q$ allows the time dependence of the surface roughness,

\begin{equation} \label{eq:INTRMSroughness}
\sigma_{rms} ^2(t)  = \int_0^{\infty}dq \frac{q}{2\pi}PSD(q,t)~~,
\end{equation}

\noindent which yields  a power law $\sigma_{rms} \sim t^ {\beta}$.  Therefore, the PSD evolution contains the complete spatial and temporal descriptions of the surface morphology evolution. Note that for EW surface dynamics, $\alpha =0, \beta =0, z = 2$, and so:
\begin{equation} \label{eq:EWRMSroughness}
\sigma_{rms} ^2(t)  = \frac{v\Omega}{8\pi a_2}ln \left(\frac{t}{t_{min}}\right)~~,
\end{equation}
\noindent where $t_{min} \sim l^2/a_2$ is suggested by Natterman and Tang, and $l$ is a minimum length scale.\cite{Nattermann1992}

\subsection{Structure factor of a rough surface}
Sinha $\emph{et al.}$ have shown \cite{Sinha1988} that within the distorted-wave Born approximation the structure factor for an isotropic surface is given by
\begin{equation}\label{eq:SF}
 			 S(q'_z,q_{||})= \frac{e^{-Re(q'_z)^2\sigma_{rms}^2}}{|q'_z|^2}\int_0^{\infty} r~dr~\left(e^{|q'_z|^2C(r)}-1\right)J_0(q_{||}r)~~,
\end{equation}

\noindent where $q'_z=k\left(\sqrt{n^2-\cos \alpha_i^2}+\sqrt{n^2-\cos \alpha_f^2}~\right)$ denotes the vertical momentum transfer in the sample, $k=2\pi/\lambda$ is the wave number, $n$ is the index of refraction, and $\sigma_{rms}$ is the rms roughness, as determined from specular reflectivity data as shown in Fig. \ref{fg:roughnessvspressure}. $C(r)$ is the surface correlation function with $C(0)=\sigma_{rms}^2$, and $J_0$ denotes the Bessel function of the first kind and zero order.

In the case of small roughness when $(|q'_z|\sigma_{rms})^2\ll1$, the integral of Eq. (\ref{eq:SF}) can be approximated by a Fourier transform of $C(r)$, which is the PSD of a rough surface. Therefore, the real-time x-ray scattering gives a direct way to monitor the surface morphology evolution during the film growth over the entire spectrum of accessible length scales.


%


\subsection{Extraction of the particle volume and relaxation coefficients from $S(q)$\label{SRAbreak}}

\begin{figure}[htbp]
	\includegraphics[width=6.5 in]{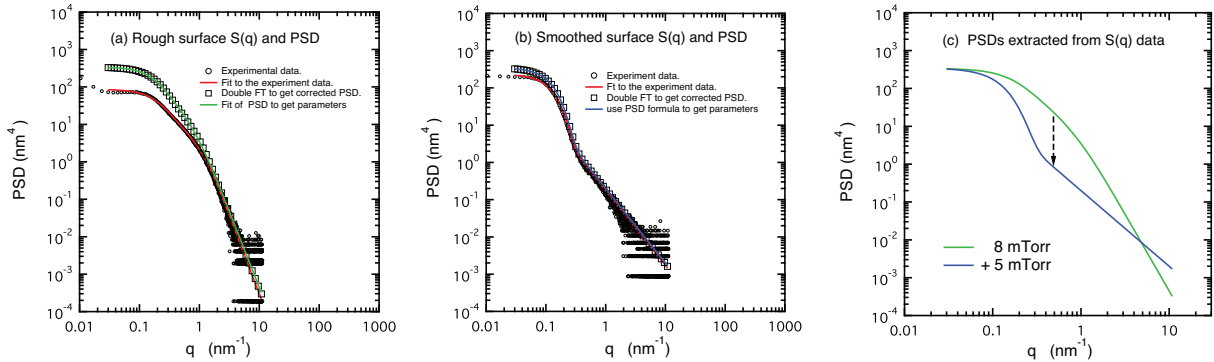}
	\caption{Extracted PSD from normalized GISAXS intensity profile for film deposition at (a) 8 mTorr on thermal SiO$_2$/Si substrate; (b) subsequent deposition at 5 mTorr on as-grown rough surface; (c)the extracted PSD evolution for the smoothing experiment, which is consistent with linear theory, i.e., see Eq. (\ref{eq:PSD}). \label{fg:smoothinganalysis}}
\end{figure}

\begin{table}[htbp]
\caption{Film parameters, particle volumes and smoothing coefficients. \label{tab:coefficient}}
\centering
\begin{tabular}{  p{0.6 in}  p{0.6 in}  p{0.6 in} p{0.6 in}  p{0.6 in}  p{0.6 in} p{0.6 in} p{0.6 in} p{0.6 in}}
\hline\hline		
  Sample & $P$ &  $t_f$ &  $\sigma_{rms}$  & $\Omega$& $a_2$  & $a_3$ & $a_4$ &    $\sqrt{a_2/a_4}$\\[1 pt]
         & (mTorr) & (nm) & (nm)& (nm$^3$)& (nm) & (nm$^2$) & (nm$^3$)  &  (nm$^{-1}$)  \\
  \hline
 I ~(A) & 8.0  & 100 & 1.60  & 3.55 & 0.34 & -0.25 & 0.42  & 0.90\\
 I ~(B)\textrm{\footnote{Subsequent~deposition~atop~layer~(A).}} & 5.0 & 220 & 0.66 & 0.038 & 0.10  &~0.00  &0.00  &~--- \\
II (A) & 8.0 & 480 & 2.84 & 3.55 & 0.12 & -0.25 & 0.42 & 0.53\\
 II (B)$^a$ & 4.0 & 600 & 0.62 & 0.020 & 0.14 &~0.00&0.00&~--- \\
 III & 5.9 & 200 & 0.39 & 0.038 & 0.12 & -0.08 & 0.08 & 1.19\\
 IV & 6.5 & 188 & 1.52 & 1.80 & 0.25 & -0.05 & 0.85& 0.54 \\
 V & 7.0 & 183 & 1.80 & 2.50 & 0.28 & -0.10 & 0.80  & 0.59\\
 VI & 8.0 & 181 & 2.10 & 3.60 & 0.35 & -0.40 & 0.85 & 0.64 \\
 VII & 10.0 & 182 & 2.57 & 6.62 & 0.50 & -0.50 & 1.00 & 0.50 \\
 VIII & 14.0 & 179 & 3.15 &N.A.\textrm{\footnote{Integration loses accuracy when $q_z \sigma_{rms} \geq 2$.}}  &N.A. &N.A.&N.A.&~--- \\
\hline\hline
\end{tabular}
\end{table}

When  $(|q'_z|\sigma_{rms})^2\ge 1$, the small roughness approximation is not valid  because we can not make the approximation  $e^{|q'_z|^2C(r)}-1 \approx |q'_z|^2C(r)$ in Eq. (\ref{eq:SF}). Then, the x-ray scattering intensity does not accurately gives the PSD of a rough surface.   In particular, we find that the scattered intensity underestimates the  PSD at the lower values of $q$. However, the PSD can be recovered  by following procedure: (i) numerically reverse Fourier transform the x-ray intensity in Eq. (\ref{eq:SF}) to get the function $F(r)\equiv e^{|q'_z|^2C(r)}-1$; (ii) normalize  $F(r)$ according to $F(r)|_{r=0}=e^{|q'_z|^2\sigma_{rms}^2}-1$, since $C(0)=\sigma_{rms}^2$; (iii) calculate the correlation function from $C(r)=ln(F(r)+1)/|q'_z|^2$; (iv) numerically Fourier transform $C(r)$ to get the corrected power spectral density.

In Fig. \ref{fg:smoothinganalysis}(a), circles represent the x-ray scattering intensity normalized with roughness $\sigma_{rms}$ for a rough surface deposited at 8 mTorr, which is also shown in Fig. \ref{fg:smoothingrealtime}(a). The red line in Fig.  \ref{fg:smoothinganalysis}(a)  is a smooth curve obtained according to Eq. (\ref{eq:PSD}) to fit the experimental data. It is used to numerically calculate the corrected PSD, which is shown in squares. The green line is the fitting of PSD to extract the smoothing coefficients, as well as the particle volume in the deposition flux. Fig. \ref{fg:smoothinganalysis}(b) shows the same data analysis for the subsequent smooth layer growth at 5 mTorr. Note that the calculation of the PSD for the overgrowth at 5 mTorr (layer B) atop the 8 mTorr layer (layer A) requires both sets of relaxation coefficients and particle volumes for layers A and B, i.e., the final curve includes the history of the first layer. The two corrected PSD curves for the  rough surface at 8 mTorr and smooth surface at 5 mTorr are also plotted in Fig. \ref{fg:smoothinganalysis}(c) for direct comparison.    The data correction and fitting procedure were repeated for several films deposited at different pressures, and the extracted smoothing coefficients and particle volume are listed in Table \ref{tab:coefficient}.

\begin{figure}[htbp]
	\includegraphics[width=2.0 in]{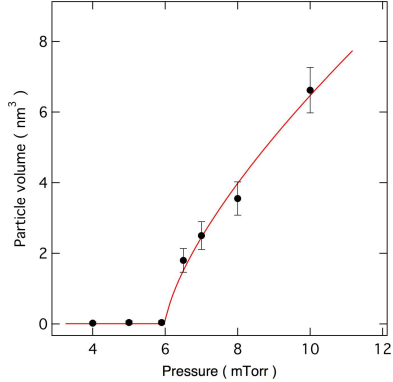}
	\caption{Particle volume produced by sputtering source extracted from GISAXS spectra. The line is a power law $(P-P_c)^m$ for the transition with an exponent of $m = 0.70 \pm 0.10$ and transition pressure at 6 mTorr. \label{fg:ParticleVolume}}
\end{figure}

The particle volume, $\Omega$ from Table \ref{tab:coefficient}, is displayed vs sputtering pressure  in Fig. \ref{fg:ParticleVolume}. Clearly, there are two main regimes separated by a sharp transition at $\sim$ 6 mTorr. The particle volume changes by two orders of magnitude as the transition boundary is crossed. A single WSi$_2$ molecule has a volume of 0.042 nm$^3$, so the particle volume above the transition pressure 6 mTorr corresponds to several hundred atoms, while below 6 mTorr the deposition flux mainly consists of single atoms and/or very small (few atoms) clusters. The particle volume is described by a power law $(P-P_c)^m$ with exponent of $m = 0.7 \pm 0.10 $. In Fig. \ref{fg:roughnessvspressure}, we have shown that $\sigma_{rms} \sim (P-P_c)^s$ with $s=0.30 \pm 0.05$, so that  $m \approx 2s$.  Comparing to Eq. (\ref{eq:EWRMSroughness}), we see that  $\sigma_{rms} \propto \sqrt \Omega$ is expected for linear theories of kinetic roughening. Evidently, $\sigma_{rms}$ is dominated by the change in particle volume, and is only affected slightly by changes in the strength of relaxation processes.

\section{Stress Evolution results  for  WSi$_2$/Si multilayers \label{multilayerresults}}

\begin{figure}[htbp]
	\includegraphics[width=3.0 in]{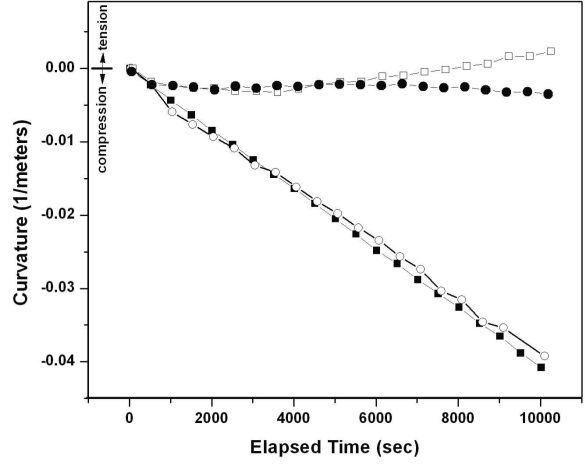}
	\caption{Curvature data plotted as a function of sputtering time for Ar plasma pressures of 2.3 mTorr (solid squares), 6 mTorr (open circles), 12 mTorr (solid circles), and 18 mTorr (open squares). Twenty bilayer periods of thickness 11.0 nm each were deposited at each pressure.  The raw data were shifted vertically to agree at the growth start. \label{fg:stressvstime}}
\end{figure}

\begin{figure}[htbp]
	\includegraphics[width=3.0 in]{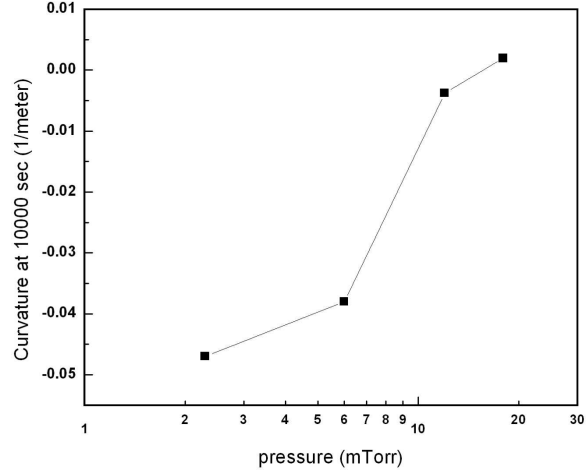}
	\caption{Curvature values extracted from Fig. $\ref{fg:stressvstime}$ at the conclusion of growth, that is, at 10000 s at which time 220 nm had been deposited, as a function of sputtering pressure.  \label{fg:stressvspressure}}
\end{figure}

$\emph{in situ}$ curvature data plotted as a function of sputtering time for different Ar pressure are shown in Fig. \ref{fg:stressvstime}. As indicated in the figure, compression produces a negative curvature (convex deformation) according to our sign convention for curvature, and vice-versa for tension. The compressive film stress develops at low sputtering pressures, while tensile stress is observed for high sputtering pressures. The curvature values at the conclusion of growth are plotted as a function of pressure in Fig. \ref{fg:stressvspressure}. It clearly shows a transition from compressive to tensile stress with increasing sputtering pressure. The transition pressure is at 6 mTorr, in good agreement with the results of the film roughness experiment. It suggests that the transition from compressive stress to tensile stress is correlated with the transition from smooth to rough film growth.

The stress transition is very similar to those reviewed in Fig.1.39 in the text by Freund and Suresh for the metals Cr, Mo, Ni, and Ta, and they are consistent with models in which tensile stress has been linked to island coalescence.\cite{Freund2003, Mayr2001} The compressive stress observed below the transition can be interpreted as a result of "atomic peening", which is due to the energetic particles striking the growing film with their high impact kinetic energy.

\section{DISCUSSION OF MECHANISMS \label{sec:discussion}}
This work combines several topics that have developed independently: the kinetics of surface evolution and dynamic scaling theory, cluster condensation in the gas phase, and pressure-dependent roughness and stress transitions of films during magnetron sputter deposition. In this section, we will discuss the mechanisms for each process.

\subsection{Plasma-gas condensation \label{sec:PGCdiscussion}}

\begin{figure}[htbp]
	\includegraphics[width=3.0 in]{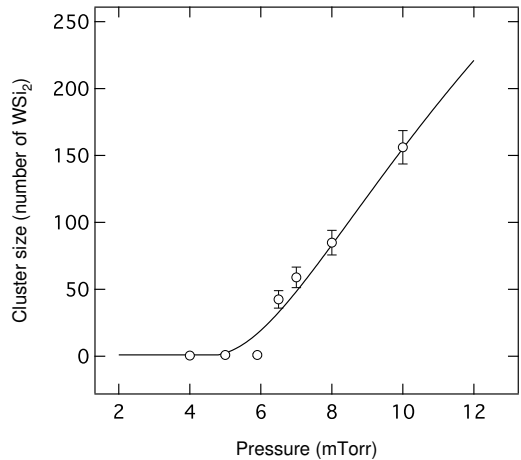}
	\caption{Model for pressure dependence of cluster growth incorporating the effects of thermalization and aggregation (line).  The parameters used are $P_c$ = 4.8 mTorr,  $v_d/n_0 = 8.5\times10^{-20}$ m$^4$/s,  and $N_{c,0}=1$.  With these parameters, the cluster size in the high pressure limit, $P \rightarrow \infty$ in Eq. (\ref{eq:aclustersize}), is found to be 833. The data points are the cluster volumes from Table \ref{tab:coefficient} divided by the volume of a single WSi$_2$ cluster.\label{fig:clustergrowthmodel}}
\end{figure}

Current models for inert gas aggregation of nanometer scale clusters postulate a nucleation step followed by cluster growth  due to atomic vapor condensation onto the newly-formed embryos.\cite{Soler1982, Hihara1998} In order for nucleation to occur, the kinetic energy of the sputtered atoms must be reduced to thermal energies via collisions with the background gas.  While plasma-gas condensation instruments are typically operated at high pressures of 200 - 300 mTorr where thermalization occurs over distances of a few millimeters,\cite{Xirouchaki2004} in low-pressure sputter deposition thermalization occurs over a distance that can be comparable to the source - sample distance.  Somekh\cite{Somekh1984} has performed Monte Carlo studies of this process and has found that the thermalization can be characterized by a pressure-distance (PD) product that gives the critical pressure at a given drift length. Their results predict that W sputtered atoms with an initial kinetic energy of 5 - 25 eV lose 90\% of their energy with a PD in the range of $\sim$ 120 Pa-mm (900 mTorr mm).  At $P=P_c$, PD$ = P_cL$, where $L$ is the distance between the sputtering source and the sample.   Given our drift length of $L=90$ mm, a critical pressure of 10 mTorr is predicted, which is in reasonably close agreement with our observation for WSi$_2$ of $P_c$ = 6 mTorr.  Therefore, our results are consistent with the point of view that the thermalization of sputtered vapor is a controlling factor in determining $P_c$ for  cluster growth, as well as the roughness and stress transitions.

The picture above implies that particles leaving the sputtering source travel ballistically towards the substrate surface until they are thermalized, and only then begin to coalesce into clusters.  At this stage, the density of the sputtered vapor must equal a density $n_0$  so that the clusters grow to a size, $N_c$ atoms per cluster.  Hihara and Sumiyama \cite{Hihara1998} have developed a model based on collision cross sections in the vapor phase to describe the growth of clusters in plasma-gas condensation instruments at high pressures.  We will adapt this model for low-pressure sputter deposition in order to approximately relate the sputtered metal vapor density $n_0$ to the particle size.  Integrating the expression given by Hihara and Sumiyama for $dN_c/dz$, where the $z$ axis is in the direction from the sputtering source to the sample, neglecting re-evaporation from clusters, we have

\begin{equation}
\label{eq:vaportdensity}
n_0 = \frac{ 3 N_c^{1/3}  v_{d}  }{ \pi r_a^2~ \bar{v}_{th}  L } 	
\end{equation}

\noindent where $\bar{v}_{th} = (8 k_B T/ \pi m_a)^{1/2}$ is the thermal velocity of the metal atoms in the vapor, $r_a$ is their atomic radius, and $m_a$ is the atomic mass.  Also, $v_{d}$ is the drift velocity of the clusters.  Note that in our calculations we take a WSi$_2$ subcluster to be the fundamental unit rather than individual atoms for simplicity.

From Table \ref{tab:coefficient}, we see that the particle size for 10 mTorr deposition is about 6.6 nm$^3$.  This corresponds to $N_c$  = 156 WSi$_2$ subclusters. The thermal velocity of a WSi$_2$ subcluster is 163 m/s,   and $r_a$ = 0.216 nm, $m_a = 3.99\times10^{-25}$ kg. Using the thermal velocity  of a fully developed cluster with $N_c$  = 156 , we have as the drift velocity $v_d = 13.0$ m/s,  which yields an estimate of  $n_0 \approx 9.7\times10^{19}$ m$^{-3}$.  This is a remarkably high vapor density given that the density of the background Ar at 10 mTorr and 300 K is $3.2\times 10^{20}$ m$^{-3}$. As a check of the consistency of this result with our experiment, we also relate this number with an estimate for the flux of sputtered particles.  Given $I$ = 100 mA  and $V$ = 500 V on the target, a target diameter of 2 in., and assuming a sputtering yield of approximately one WSi$_2$ per incident ion, we estimate $J \approx 3.1 \times 10^{20}$ m$^{-2}$-s$^{-1}$.  Combining this with the relation $J = n_0 v_{d}$ and substituting into Eq. (\ref{eq:vaportdensity}), we find that $n_0 \approx 4.8 \times 10^{19}$ m$^{-3}$, which is close to the $a ~priori$ estimate obtained from Eq. (\ref{eq:vaportdensity}) alone.   Therefore, although there is significant uncertainty in the parameter $v_{d }$ on the order of a factor of 2,  these estimates suggest that a sputtered metal vapor density of $10^{19}$--$10^{20}$ m$^{-3}$ is required for production of clusters in sputter deposition. Since the corresponding telltale effects of tensile stress and roughness have been reported frequently in the published literature  on sputter deposition of metal films, a very high sputtered vapor density must be inherent to the magnetron sputtering process.

Equation (\ref{eq:vaportdensity}) does not predict any significant pressure dependence. However, it is possible to construct a model combining the effects of thermalization and cluster growth by substituting an effective drift length $L_{eff}$ for the total drift length $L$  where $L_{eff}$ is the net drift length over which clusters form, i.e., $L_{eff} = L - L_{th}$. Here, $L_{th}$ is the thermalization length  $L_{th} = P_cL/P$, so that  $L_{eff} = L(1-P_c/P)$ . We note that $L_{eff}$ should be zero at $P = P_c$.    Rearranging Eq. (\ref{eq:vaportdensity}), substituting $L_{eff}$ for $L$, and explicitly including the embryo cluster size $N_{c,0}$, we have

\begin{equation}\label{eq:aclustersize}
N_c \approx
\begin{cases}
N_{c,0} & \text{for}  P\le P_c~,\\
\left[   (\pi r_a^2 n_0 \bar{v}_{th} L / 3 v_d )     \left(1-\frac{P_c}{P} \right)   + N_{c,0}^{1/3}  \right]^3 & \text{for } P > P_c~.
\end{cases}
\end{equation}

This model successfully predicts a strong pressure dependence near $P_c$, as we show in Fig. \ref{fig:clustergrowthmodel}.  The curve matches the experimental values well, except for within $\sim$ 1 mTorr of the transition.  This relatively good agreement suggests that the model captures the essential mechanisms responsible for the cluster transition.  We emphasize that although a  model can be constructed based on the pressure dependence of  particle impact smoothing without any cluster aggregation, such a  model will not resemble our particle volume vs pressure data because  the smoothing processes vary continuously with particle energy, i.e., they do not exhibit a threshold at a definite pressure.   Finally, we comment that the exponent used in the fit of Fig. \ref{fg:ParticleVolume} is not likely to be a universal exponent.  Rather, the detailed shape of the cluster size vs pressure curve appears to be the result of at least two processes operating  in concert, and thus may change under different experimental conditions.

\subsection{Kinetic roughening}

As discussed in Sec. \ref{sec:dynamics}, kinetic roughening is the result of the competition between the inherent noise in the growth process due to the nonuniform nature of the incoming flux and surface relaxation effects. Each smoothing mechanism affects the morphology differently and leaves a different signature in the PSD of the growing surface.

The roughening in the low-pressure regime is exemplified by the GISAXS spectra shown in Fig. \ref{fg:psdvstimeattwopresures}(b), and the smoothing layer deposited on a rough first layer shown in Fig. \ref{fg:smoothingrealtime}(a).  Strong smoothing is observed, and   both curves exhibit a crossover  to stable  $q^{-2}$ behavior above a cutoff  wave number $q_c$ that shifts toward lower $q$ with time. The fact that this behavior is independent of the starting surface is consistent with kinetic roughening, as can be seen in Eqs. (\ref{eq:PSD}) - ( \ref{eq:PSDinfty}). The classic linear model involving only the $a_2$ coefficient in Eqs. (\ref{eq:FTLangevin}) and (\ref{eq:bq}),  corresponding to a $\nabla ^2 h$ term in real space, was originally described by Edwards and Wilkinson for the case of particle sedimentation.\cite{EW1982}   EW behavior has been observed before in sputter deposition at low pressures.\cite{Salditt1994, Salditt1996} More recently, this term has been suggested as being related to energetic impact  induced downhill relaxation, which is expected to occur when energetic particles reach the growth surface with energies exceeding $\sim$20 eV.\cite{Moseler2005, Carter2001}

The observed GISAXS spectra are more complicated in the higher pressure range.  However, the detailed fitting results give some insight into the mechanisms.  The $a_2$ term listed in Table \ref{tab:coefficient} is related to the EW ($\nabla ^2 h$) term, which results in the $q^{-2}$ behavior  at $q>q_c$.  At first sight, it is somewhat curious that the $a_2$ coefficient  increases significantly for pressures above $P_c$ because we would expect that impact-induced smoothing would be significantly suppressed above the thermalization pressure.  However, relaxation of clusters into nearby hollows in the surface is also described by an EW-type term.  Thus, we interpret the increase in the $a_2$ coefficient as the appearance of a new relaxation process related to cluster relaxation on the growth surface.  The monotonic increase in $a_2$ with pressure is attributed to the increasing particle size.  The $a_4$ term can be interpreted similarly as a shorter length scale process related to the sintering of clusters.  For example, processes such as viscous flow can give a $\nabla^4 h$ term.\cite{Umbach2001}    The  $a_4$ term can also be related to Mullins surface diffusion,\cite{Mullins1957}  although surface diffusion may be negligible at room temperature.   The $a_3$ term  has a negative value, which  shows that it is not a relaxation process.  Rather it could indicate a surface instability process such as strain-driven diffusion.\cite{Asaro1972} However, we emphasize that we have not found evidence for unstable surfaces under any of the experimental conditions, indicating that $b(q) > 0$ for all $q$.

\subsection{Stress transition}

The stress transition is a particularly striking effect that has been demonstrated for a variety of sputter-deposited films.\cite{Freund2003, Hoffman1977, Hoffman1980}    In polycrystalline films tensile stress observed at pressures above the transition has been suggested to be due to closing the small gaps between islands to form grain boundaries, which ultimately merge into a continuous film.\cite{Doljack1972, Nix1989, Nix1999}

A similar concept has been suggested as a way to explain tensile stress in amorphous films, which can form cellular structures characterized by grooves at the boundaries between hillocks.\cite{Mayr2001, Tang1990, Salditt1996}     The growth of the cellular structures is driven by the deepening of grooves due to shadowing effects, which can contribute to tensile strain by closing of the grooves.  However, we point out that the coarsening behavior observed for these layers is distinct from kinetic roughening because  larger features increase in size at the expense of smaller features and hence there is no steady-state power spectrum over any range of length scales.\cite{Tang1990}  Therefore, this type of coarsening is only relevant to a later stage of the film deposition, while our observations are for the early stage when  shadowing effects are absent and the groove stress mechanism does not play a dominant role in producing stress.  The fact that the stress transition in our data is coincident with the formation of clusters strongly suggests that the clusters directly produce the stress by sintering  and elimination of voids between particles, which can be very fast for nanometer scale particles even at room temperature.\cite{zhu1996}  This shrinkage  effect is well known in powder metallurgy and is due to that fact that as particles fuse together their centers also move closer together.\cite{German1996}

To conclude this discussion, we point out that the observation of clusters in magnetron sputtering suggests many opportunities for tailoring the properties of thin films.  Aside from the control of the film growth process by varying the background pressure, other parameters, particularly the sputtering power and the type of sputtering gas are expected to have a strong effect on cluster formation and in the properties of films thus produced.

\section{CONCLUSIONS}
In conclusion, we have developed a comprehensive model that encompasses several effects occurring in sputter-deposited thin films and multilayers. Specifically, we have shown that the roughening transition and the stress transition both arise from aggregation in the sputtering plasma as pressure increases above a threshold value.

%



\begin{acknowledgments}

The authors acknowledge Nathalie Bouet, Lin Yang, Christie Nelson, D. Peter Siddons,  and Tony Kuczewski, for experimental assistance with the work done at the NSLS X21 beamline, and Lihua Zhang for experimental assistance with TEM at the Center for Functional Nanomaterials. The authors also acknowledge the support for K. MacArthur Êprovided by Prof. A. Genis at ÊNorthern Illinois University.  This material is based on work supported by the U.S. Department of Energy under Grant No. DE-FG02-07ER46380. Use of the National Synchrotron Light Source and the Center for Functional Nanomaterials, Brookhaven National Laboratory, was supported by the U.S. Department of Energy, Office of Science, Office of Basic Energy Sciences, under Contract No. DE-AC02-98CH10886.  Work at  Argonne was performed  under Department of Energy contract No. DE-AC02-06CH11357. \\

\end{acknowledgments}


\bibliography{Lan_Sputter_Deposition_V14}

\end{document}